\documentclass[twocolumn,english,aps,prb,showpacs,byrevtex,amsmath,amssymb,superscriptaddress]{revtex4-1}
\usepackage[T1]{fontenc}
\usepackage[latin9]{inputenc}
\usepackage{lineno}
\usepackage{textcomp}
\usepackage{amsmath}
\usepackage{graphicx}
\usepackage{multirow}
\usepackage{ bbold }
\usepackage{lipsum}
\usepackage{array}
\newcolumntype{P}[1]{>{\centering\arraybackslash}p{#1}}

\makeatletter
\usepackage{lipsum}
\usepackage{epsfig}

\usepackage[colorlinks,linkcolor=blue]{hyperref}
\usepackage[nameinlink,capitalise]{cleveref}
\usepackage{subfigure}
\usepackage{color}
\usepackage{physics}
\usepackage[title]{appendix}

\definecolor{Red}{rgb}{1,0,0}
\definecolor{Blu}{rgb}{0,0,1}
\definecolor{Green}{rgb}{0,1,0}

\makeatother
\usepackage{amssymb}
\usepackage{babel}

\usepackage{tikz,xcolor,hyperref}
\definecolor{lime}{HTML}{A6CE39}
\DeclareRobustCommand{\orcidicon}{%
	\begin{tikzpicture}
	\draw[lime, fill=lime] (0,0)
	circle [radius=0.16]
	node[white] {{\fontfamily{qag}\selectfont \tiny ID}};
	\draw[white, fill=white] (-0.0625,0.095)
	circle [radius=0.007];
	\end{tikzpicture}
	\hspace{-2mm}
}

\foreach \x in {A, ..., Z}{%
	\expandafter\xdef\csname orcid\x\endcsname{\noexpand\href{https://orcid.org/\csname orcidauthor\x\endcsname}{\noexpand\orcidicon}}
}

\usepackage{amsmath}
\usepackage[TS1,T1]{fontenc}
\usepackage{textcomp}

\begin{document}
\title{Sign competing sources of Berry curvature and \\ anomalous Hall conductance humps in topological ferromagnets} 
\author{Wojciech Brzezicki\orcidB}
\affiliation{Institute of Theoretical Physics, Jagiellonian University, ulica S. \L{}ojasiewicza 11, PL-30348 Krak\'ow, Poland} 
\affiliation{International Research Centre MagTop, Institute of Physics, Polish Academy of Sciences, Aleja Lotnik\'ow 32/46, PL-02668 Warsaw, Poland}

\author{Carmine Autieri\orcidA} 
\email{autieri@magtop.ifpan.edu.pl}
\affiliation{International Research Centre MagTop, Institute of Physics, Polish Academy of Sciences, Aleja Lotnik\'ow 32/46, PL-02668 Warsaw, Poland}
\affiliation{SPIN-CNR, c/o Universit\`a di Salerno, IT-84084 Fisciano (SA), Italy}

\author{Mario Cuoco\orcidC}
\affiliation{SPIN-CNR, c/o Universit\`a di Salerno, IT-84084 Fisciano (SA), Italy}

\begin{abstract}
\textbf{\normalsize 
The use of Berry-phase concepts has established a strong link between the anomalous Hall effect (AHE) and the topological character of the Hall currents. 
However, the occurrence of sign competition in the Berry curvature often hinders the topological origin of the observed anomalous Hall effects.
Here, we study a two-dimensional topological ferromagnet with coupled spin and orbital degrees of freedom to assess the anomalous Hall effects in the presence of sign-competing sources of Berry curvature.  
We show that itinerant topological ferromagnets can generally lead to topological metallic bands marked by a non-zero Chern number. We find that the resulting Berry curvature exhibits a characteristic anisotropic profile with a non-monotonous angular dependence. The sign change of the intrinsic contribution to the anomalous Hall conductance can occur together with topological transitions or be driven by the population imbalance of the topological bands. 
The breaking of the inversion symmetry introduces the orbital Rashba coupling in the system. The interplay between the orbital Rashba and sign competing sources of Berry curvature leads to anomalies in the anomalous Hall conductance at values of magnetic fields for which the magnetization switches its orientation. The humps in topological ferromagnets arise when the anomalous Hall conductivity is small in absolute value and they can be detected only far from half-filling. 
This study could be relevant for the family of the topological 2D ferromagnets as well as Weyl ferromagnets, and can particularly account for the variety of unconventional behaviors observed in ultrathin films of SrRuO$_3$.}   
\end{abstract}
\maketitle


\vskip 0.3cm
\textbf{\Large Introduction}
\vskip 0.3cm

In the last decades, a lot of attention has been devoted to topological phases in condensed matter.\cite{Nagaosa2020-si,Singh2023-dy} 
Apart from its fundamental relevance, the focus on materials with topologically protected quasiparticles in momentum space (e.g., Dirac or Weyl semimetals) or those with non-trivial
real-space spin textures, together with the use of degrees of freedom alternative to electric charge, as the electron spin, is rapidly setting the path for a topological view of spintronics and spin-orbitronic effects. One of the central physical quantities for identifying and exploiting the topological behavior of quantum materials is the Berry curvature, which is known to capture the essence of quantum Hall and quantum spin Hall effects and describes pseudo-magnetic fields arising from the quantum geometric properties of electron wave functions.
It is generally expected that Berry curvatures with opposite signs can contribute to the Fermi level because the total Chern number evaluated over all the bands in the Brillouin zone must be zero. Such sign competition, if rooted in topologically non-trivial bands at the Fermi level, will have a direct impact on the behavior of the anomalous Hall conductivity and can hinder the topological origin of the Hall currents.\cite{RevModPhys.82.1959,10.21468/SciPostPhysLectNotes.51} 

An emblematic case of sign-tunable anomalous Hall conductance\cite{vanThiel2021coupling} is given by the SrRuO$_3$ compound, where the anomalous Hall conductivity is positive at high temperatures and becomes negative at low temperatures. Three-dimensional SrRuO$_3$ is a Weyl ferromagnetic metal\cite{Itoh2016-na, PhysRevB.88.125110,reviewSIO} widely used as an oxide electrode in functional heterostructures and for spintronics and spin-orbitronics applications\cite{Koster2012,Trier2022,Kimbell2022,Gu2022}. 
SrRuO$_3$ hosts nonlinear and nonreciprocal transport effects\cite{PhysRevX.14.011022,sha2025signreversalberrycurvature}.
Over the past years, ultrathin and bulk SRO films have been at the center of intense research for the anomalies observed in magnetic and transport behavior \cite{Takiguchi2020, Kan2020, Groenendijk2020, Wang2020a, vanthiel2020coupling, Li2020, Wang2018, Fang, Ohuchi2018, Kan2018, Matsuno2016,vanThiel2021coupling} as well as anomalous Hall response governed by momentum space topology. \cite{Fang,Groenendijk2020,vanThiel2021coupling,Sohn2021} 
At the same time, structural modifications can tune the AHE\cite{Ziese2019-ci}. 


SrRuO$_3$ hosts humps in the AHE hysteresis. The humps in the SrRuO$_3$ system are sensitive to ferroelectric proximity\cite{Yao2022-zx,Wang2018}, asymmetric slabs\cite{Ziese2018-un} or electric field\cite{PhysRevB.96.214422,Ohuchi2018}, therefore, the breaking of the inversion symmetry definitely plays a role. 
The humps in the AHE are usually associated with the topological Hall effect described by the presence of real-space topological objects as skyrmions\cite{WANG2022100971,PhysRevX.15.011054} or other non-topological complex magnetic structures\cite{PhysRevApplied.17.064015, dao2022halleffectinducedtopologically}. These complex magnetic structures often require the breaking of the inversion symmetry.
However, several authors have proposed that the humps in SrRuO$_3$ could be due to band structure and therefore to k-space effects\cite{Malsch2020-hw,PhysRevB.102.220406,Tian2021-mw,PhysRevMaterials.4.014401,Roy2023-df}. 
A tunability of the humps is present in the case of Ru vacancies\cite{Kan2018-bn} or dirty-regime\cite{PhysRevMaterials.7.054406}. From the necessity to stay in the dirty regime, the authors conclude that the Weyl topology is not enough to generate the humps\cite{PhysRevMaterials.7.054406}. 
For the humps generated by the k-space, different approaches mimicking the topological Hall effect in SrRuO$_3$ have been proposed, such as the presence of two Hall channels due to inhomogeneity\cite{SOHN2020186}. In the earlier papers, this Hall channels approach was rather phenomenological; indeed, this two-channel AHC assumes 2 non-interacting AHC loops. However, we cannot disentangle the Berry curvature associated with different layers or regions of the real space in two non-interacting loops due to the strong hybridization between the different regions to create a unique band structure\cite{vanThiel2021coupling}. More detailed models have been proposed where it was highlighted that the relevance of the multidomains in Weyl ferromagnets with sign competing sources of Berry curvature\cite{sabri2025topologicalhalllikebehaviormultidomain,ard2025boundsanomaliesinhomogeneousanomalous}. 
Among the experimental techniques to disentangle the hump-shape Hall effects for distinguishing between k-space and real space origin, it was proposed to study the AHE angular dependence\cite{Lim2025-ia,Esser2021angular,PhysRevResearch.3.023232} or minor loops\cite{Tai2022-lg}.

Going beyond SrRuO$_3$, the presence of humps without clear detection of skyrmions is also present in other compounds\cite{PhysRevX.10.011012,Krempasky2023,Fujisawa2023-qb,Chi2023-ri,Shen2023-hr}, while the sign change of the AHE is found in several ferromagnetic Weyl semimetals\cite{PhysRevB.107.085102,Fujisawa2023-qb,D4MH01875C}.

In this paper, we investigate the Berry phase of a 2D topological ferromagnet without and with inversion symmetry. A topological ferromagnetic metal is characterized by bands with a non-trivial Chern number. This implies that, even in the absence of a full gap in the energy spectrum for a given electron filling, there are sources of Berry curvature. To achieve competing signs of the Berry curvature, a centrosymmetric 2D ferromagnetic topological model or a Weyl ferromagnetic model would be enough. 
However, the topology alone is not sufficient to produce humps; in addition, the orbital Rashba effect arising from the breaking of inversion symmetry must be included.
The interplay between the competing sign-change of the Berry phase and the orbital-Rashba from the breaking of inversion symmetry can produce humps in the AHE within a collinear magnetic phase.
The manuscript is organized as follows: in the next section, we describe our results, separating the model definition, the sign-competing analysis, and the description of the humps. In the final section, we discuss the consequences of these results and present our conclusions.

\vskip 0.3cm
\textbf{\Large Results}
\vskip 0.3cm

\textbf{Model of the 2D topological ferromagnet}
\vskip 0.2cm

We consider a 2D tight-binding model for $t_{2g}$ orbitals, assuming a square lattice geometry where the inversion symmetry is broken.
The assumption of the 2D square lattice has been made for convenience and clarity. Additionally, considering applications to layered oxides such as the SrRuO$_3$ and similar perovskite compounds, the 2D square lattice serves as a relevant case of study.
Nevertheless, we anticipate that comparable qualitative results can be achieved across various lattice geometries, as long as there exists a local manifold in the orbital space characterized by an angular momentum with amplitude one ($L=1$), and the orbital structure of the hopping connectivity can lead to bands with nonvanishing Chern number within the inversion-symmetric regime.
The examined 2D Hamiltonian in the $k$-space can be expressed as
\begin{eqnarray}
{\cal H}_{k} & = & -2t\cos k_{x}L_{x}^{2}-2t\cos k_{y}L_{y}^{2}\nonumber \\
 & - & t_{d}\sin k_{x}\sin k_{y}\left(L_{x}L_{y}+L_{y}L_{x}\right)\nonumber \\
 & + & \lambda_{R}\left(L_{x}\sin k_{y}-L_{y}\sin k_{x}\right)+\lambda\vec{L} \cdot \vec{\sigma}-\tfrac{\vec{h}}{2}\cdot \vec{\sigma}\label{eq:Ham}
\end{eqnarray}
where $L_{\alpha}$ are the components of the atomic angular momentum $L=1$ which have
the matrix elements $\left(L_{\alpha}\right)_{\beta\gamma}=-i\varepsilon_{\alpha\beta\gamma}$
in the $t_{2g}$ manifold of the ($d_{xz},d_{yz},d_{xy}$) orbitals, $\sigma_{\alpha}$ are the Pauli
matrices describing the spin of electrons. The nearest-neighbor hopping
amplitude is given by $t$, next-nearest-neighbor hopping by $t_{d}$
(diagonal hopping), $\lambda$ is the spin-orbit coupling and $\lambda_{R}$ is the
orbital Rashba coupling for the nearest-neighbor bonds. The orbital Rashba coupling originates from the absence of inversion symmetry. It shares the same form as the spin Rashba coupling but connects the crystal momentum to the orbital angular momentum components \cite{Park2011,Park2012,Kim2013,Mercaldo2020}. 
\\
We employ an effective Zeeman
field $\vec{h}$ to describe the occurrence of a finite magnetization in the ferromagnetic phase along the direction set by the field. 
If $\lambda_{R}=0$ and $\vec{h}=\left(0,0,h_{z}\right)$ then ${\cal H}_{k}$
commutes with the spin-orbital parity operator 
\begin{equation}
{\cal P}=\left(1-2L_{z}^{2}\right)\sigma_{z},
\end{equation}
which becomes a mirror reflection with respect to the $(001)$ plane in the case of a multilayer system. In the eigenbasis of ${\cal P}$ the
Hamiltonian becomes block-diagonal with 3x3 diagonal blocks given by 
\begin{eqnarray}
{\cal H}_{k}^{+} & = & -2t\cos k_{x}\left(1-\tilde{L}_{x}^{2}\right)-2t\cos k_{y}\left(1-\tilde{L}_{y}^{2}\right)\nonumber \\
 & - & t_{d}\sin k_{x}\sin k_{y}\left(\tilde{L}_{x}\tilde{L}_{y}+\tilde{L}_{y}\tilde{L}_{x}\right) \\
 & + & \lambda\left(\tilde{L}_{y}\tilde{L}_{z}+\tilde{L}_{z}\tilde{L}_{y}-\tilde{L}_{y}+\tilde{L}_{z}\right)+\tfrac{h_{z}}{2}\left(2\tilde{L}_{z}^{2}-1\right)\nonumber ,
\end{eqnarray}
and

\begin{eqnarray}
{\cal H}_{k}^{-} & = & -2t\cos k_{x}\left(1-\tilde{L}_{y}^{2}\right)-2t\cos k_{y}\left(1-\tilde{L}_{z}^{2}\right)\nonumber \\
 & - & t_{d}\sin k_{x}\sin k_{y}\left(\tilde{L}_{y}\tilde{L}_{z}+\tilde{L}_{z}\tilde{L}_{y}\right) \\
 & - & \lambda\left(\tilde{L}_{z}\tilde{L}_{x}+\tilde{L}_{x}\tilde{L}_{z}-\tilde{L}_{z}-\tilde{L}_{x}\right)-\tfrac{h_{z}}{2}\left(2\tilde{L}_{x}^{2}-1\right)\nonumber ,
\end{eqnarray}
with $\tilde{L}_{\alpha}$ being $\tilde{L}=1$ angular momentum operators, but {\it not the same} as in Eq. (\ref{eq:Ham}).
The two blocks ${\cal H}_{k}^{\pm}$ are related by time-reversal symmetry ${\cal T}:(\tilde{L}_{x},\tilde{L}_{y},\tilde{L}_{z})\to(\tilde{L}_{y},-\tilde{L}_{z},\tilde{L}_{x})$
and $h_{z}\to-h_{z}$. The Berry curvature for both $3\times3$ Hamiltonians
can be obtained using the SU($3$) Gell-Mann matrices formalism \cite{Bar12,Graf2021} (see also Methods). 

We note that for each block ${\cal H}_{k}^{\pm}$ the term $t_d$ is relevant in inducing a nontrivial Chern number in the electronic bands. This aspect can be deduced by the form of the Berry curvature and the fact that the Hamiltonian, apart from terms that are linear in the orbital moment, $L_\alpha$, contains terms with nontrivial quadrupole structure, such as $({L}_{x}{L}_{y}+{L}_{y}{L}_{x}$) or similar combinations (see Methods). These terms can generally occur when orbitals with different mirror parity hybridize and thus are not limited to the case of the 2D square lattice \cite{donofrio2025,Mercaldo2023}. The momentum-dependent form factor yields an orbital quadrupole texture profile that influences the behavior of the Berry curvature.
\\
As demonstrated above, the Hamiltonian can be separated into two blocks, which helps to understand the individual contributions to the Berry curvature. Without orbital Rashba coupling, the electronic states within each block -- characterized by different spin-orbital parities -- do not mix. Due to the time-reversal symmetry correspondence, these states can carry the same nontrivial topological number. However, the exchange field causes an energy splitting between them, pushing the bands with positive parity to higher energies compared to those with negative parity. Consequently, the occupancy of bands in each block differs. This imbalance in turn leads to a nonzero total Berry curvature for the occupied states below the Fermi level. When the orbital Rashba coupling is introduced, states of opposite parity become coupled. As we will see in the next Section when examining the anomalous Hall effects, the resulting behavior of the Berry curvature is determined by the combined effects of the exchange field, dictating the occupancy, and the mixing of states with different parity through the orbital Rashba interaction, resulting in a rich Berry curvature and Hall response.

It is interesting to observe that in the limit of a high exchange field, one can assume that the spin $\vec{\sigma}$
is fully aligned to $\vec{h}$, hence the Hamiltonian (\ref{eq:Ham}) becomes

\begin{eqnarray}
{\cal H}_{k}^{{\rm hf}} & = & -2t\cos k_{x}L_{x}^{2}-2t\cos k_{y}L_{y}^{2}\nonumber \\
 & + & t_{d}\sin k_{x}\sin k_{y}\left(L_{x}L_{y}+L_{y}L_{x}\right)\nonumber \\
 & + & \lambda_{R}\left(L_{x}\sin k_{y}+L_{y}\sin k_{x}\right)+\lambda\vec{\eta}\vec{L},
\end{eqnarray}
where the field enters as an anisotropy of the SOC by
\begin{equation}
\vec{\eta}=\frac{1}{\left|\vec{h}\right|}\left(-h_{x},h_{y},h_{z}\right),
\end{equation}
and where a constant energy shift of $\frac{1}{2}|\vec{h}|$
is omitted. This effective Hamiltonian has again a $3\times3$ structure whose Berry curvature can be again studied
using the Gell-Mann matrices formalism \cite{Bar12}. 
\\
In the parity ${\cal P}$
symmetric case of $\lambda_{R}=0$ and $\vec{h}=\left(0,0,h_{z}\right)$
Hamiltonian ${\cal H}_{k}^{{\rm hf}}$ commutes with ${\cal P}^{{\rm hf}}=\left(1-2L_{z}^{2}\right)$.
In the eigenbasis of ${\cal P}^{{\rm hf}}$ Hamiltonian ${\cal H}_{k}^{{\rm hf}}$
splits into $2\times2$ block and $1\times1$ diagonal blocks. The trivial $1\times1$ block 
does not contribute any Berry curvature
to the Hall conductance $\sigma_{xy}$, whereas the $2\times2$ block contributes
opposite Berry curvatures according to the SU(2) Pauli matrices formalism
\cite{Bar12}. The combined ${\cal I}{\cal T}{\cal P}$ symmetry takes
in this basis form of ${\cal K}\left(1-2L_{z}^{2}\right)$
(where ${\cal K}$ is complex conjugation) that satisfies $({\cal I}{\cal T}{\cal P}){\cal H}_{k}^{{\rm hf}}({\cal I}{\cal T}{\cal P})^{-1}={\cal H}_{k}^{{\rm hf}}$
if only $h_{z}=0$, so non-vanishing Berry curvature is possible only for $h_{z}\not=0$. This happens because we can use the eigenbasis of ${\cal I}{\cal T}{\cal P}$ to make the Hamiltonian purely real and this kills any Berry curvature for any given $k$-point. The same is true for the initial Hamiltonian (\ref{eq:Ham}). 

\begin{table}
\begin{tabular}{|c|c|c|c|c|c|}
\hline 
 & ${\cal I}$ & ${\cal T}$ & ${\cal P}$ & ${\cal T}{\cal P}$ & ${\cal I}{\cal T}{\cal P}$\tabularnewline
\hline 
\hline 
$h_{x}$ & $1$ & $0$ & $0$ & $1$ & $1$\tabularnewline
\hline 
$h_{y}$ & $1$ & $0$ & $0$ & $1$ & $1$\tabularnewline
\hline 
$h_{z}$ & $1$ & $0$ & $1$ & $0$ & $0$\tabularnewline
\hline 
$\lambda_{R}$ & $0$ & $1$ & $0$ & $0$ & $1$\tabularnewline
\hline 
\end{tabular}\caption{Summary of the symmetry properties of the model (\ref{eq:Ham}) in
presence of terms proportional to $h_{x}$, $h_{y}$, $h_{z}$ and
$\lambda_{R}$. $0/1$ means broken/preserved symmetry. Basic symmetries
are: inversion ${\cal I}=1$, time-reversal ${\cal T}={\cal K}\sigma_{y}$
and spin-orbital parity ${\cal P}=\left(1-2L_{z}^{2}\right)\sigma_{z}$.
Breaking of combined ${\cal I}{\cal T}{\cal P}$ symmetry is a necessary
condition of having non-vanishing Berry curvature and $\sigma_{xy}$. }
\end{table}

\vskip 0.3cm
\textbf{Sign competing sources of Berry curvature \\ and Chern numbers}
\vskip 0.2cm

\begin{figure}
\includegraphics[width=1.0\columnwidth]{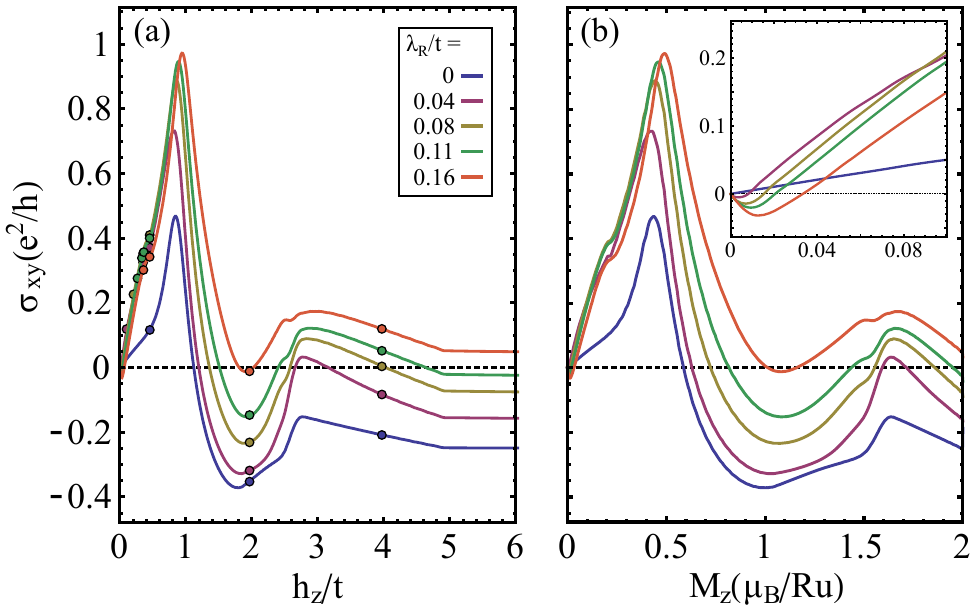}
\caption{Anomalous Hall conductivity $\sigma_{xy}$ as a function of: (a) effective
Zeeman field $h_{z}$ and (b) magnetization $M_{z}$ (region around
$M_{z}=0$ zoomed in the inset). The parameters are $t_{d}=-0.16t$,
$\lambda=-0.2t$ and $\lambda_{R}/t=0,\,0.04,\,0.08,\,0.11,\,0.16$
for decreasing value of high-field $\sigma_{xy}$, respectively. 
The colored dots in (a) indicate the reshuffling of the Chern  
numbers of the bands.
The temperature is fixed at $T=0.01t$. Here, we refer to the magnetization per Ru site, assuming that it can be applied to the SrRuO$_3$ compound and similar oxides. }\label{fig1}
\end{figure}

Using the numerical approach reported in the Methods Section, we calculate the anomalous Hall conductivity $\sigma_{xy}$ as a function of $h_{z}$ and as a function of the magnetization M$_z$ value, scanning different values of the orbital Rashba $\lambda_{R}$. 
The magnetization along the z-axis M$_z$ is defined as: 
\begin{eqnarray}
M_z=\frac{1}{V_{BZ}}\sum_{\vec{k}}\sum_{n=1}^6\Theta(-E^{(n)}_{\vec{k}}) \langle\psi_{\vec{k}}^{(n)}|\sigma_z|\psi_{\vec{k}}^{(n)}\rangle 
\end{eqnarray}
where $E^{(n)}$ and $|\psi_{\vec{k}}^{(n)}\rangle$ are eigenvalues and eigenvectors of Hamiltonian (\ref{eq:Ham}) and $\Theta$ is a step function.
The easy axis of the system is aligned along the z-axis.
For the analysis, we assume a hopping parameter $t_d=0.16 t$ and a spin-orbit coupling strength of $\lambda=0.2 t$, where all quantities are expressed in units of the nearest-neighbor hopping $t$. Then, we vary the amplitude of the orbital Rashba coupling to include the effects of inversion and mirror symmetry breaking that arise in an ultrathin system. Although we aim to maintain a general approach, these parameter values are consistent with those obtained through Wannier projection for SrRuO$_3$  
\cite{Groenendijk2020,vanThiel2021coupling}.
We note that the range of variation of the orbital Rashba coupling is also compatible with that obtained in perovskite materials due to the confinement potential along the z-axis\cite{Groenendijk2020,vanThiel2021coupling}.
\\
The behaviors of the AH conductance in terms of the exchange field and of the magnetization are reported in Fig. \ref{fig1}(a) and Fig. \ref{fig1}(b), respectively.
From Fig. \ref{fig1}(a), we can observe how $\sigma_{xy}$ is always zero in the limit of zero $h_{z}$ since the system, apart from being time-reversal, becomes also symmetric with respect to time and twofold rotation around the z-axis or equivalently the combination of time with inversion and spin-orbital parity. By increasing $h_{z}$, we observe a non-monotonic behavior in $\sigma_{xy}$, which becomes strongly positive at $h_{z}$=t to turns negative at around $h_{z}$=2t for most of the values of the orbital Rashba $\lambda_{R}$. We can observe how the 2D topological ferromagnets can host competing sources of Berry curvature with positive and negative values as a function of the strength of the magnetization. For the largest investigated value of $\lambda_{R}$=0.16t, the value of $\sigma_{xy}$ is always positive.
The black dots in Fig. \ref{fig1}(a) represent topological phase transitions where the bands change their Chern number, even if the system is still in a metallic regime and it cannot host a quantized Chern number. These phase transitions are characterized by the appearance of Weyl points in the extended $(\vec{k},h_z)$ three-dimensional space.
From the point of view of AHE, we note that they produce changes in the first and second derivatives of $\sigma_{xy}$ as a function of $h_{z}$.
A similar trend is observed when we replace $h_{z}$ with the self-consistent magnetization $M_z$. 
Since the spin is entirely along the z-axis, in this subsection, the magnitude of the magnetization $M$ coincides with its z-component M$_z$. In Fig. \ref{fig1}(b), $\sigma_{xy}$ also exhibits the presence of competing sources of Berry curvature as a function of the strength of the magnetism.

\begin{figure}
\includegraphics[width=1\columnwidth]{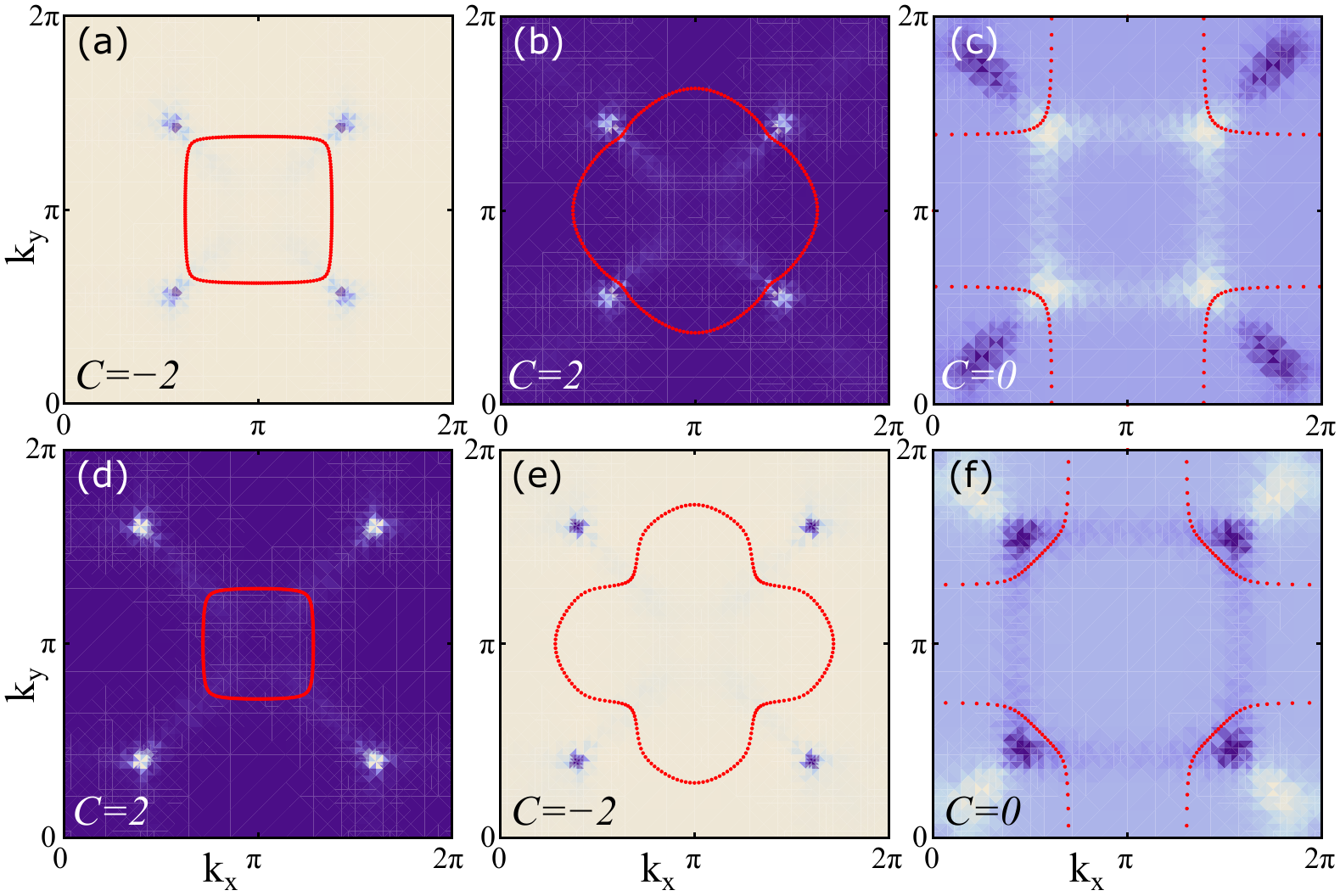}
\caption{Density plots of Berry curvature $\Omega$ and Fermi surfaces (indicated
with red dots) for all $6$ bands in the ${\cal P}$-symmetric case
of $\lambda_{R}=0$ and $\vec{h}=\left(0,0,0.5t\right)$. The bands
in (a-c) belong to the subspace ${\cal P}=1$ and are ordered with
increasing energy; the bands in (d-f) belong to the subspace ${\cal P}=-1$
and are ordered in the same way. Point $\vec{k}=\left(0,0\right)$
belongs to occupied states, the Chern numbers ${\cal C}$ of individual bands are indicated. \label{fig2}}
\end{figure}

To gain some insight into the Berry curvature, we report the Berry curvature $\Omega$ on top of the Fermi surfaces in the 2D Brillouin zone. 
The t$_{2g}$ manifold is made up of six electronic bands; for the input parameters reported, all bands contributed to the Fermi surface. The density plots of the Berry curvature are reported in Fig. \ref{fig2}. The bands are reported with increasing energy; therefore, the bands in \ref{fig2}(a) and \ref{fig2}(d) have similar character but opposite values of $S_z$, the same for the bands \ref{fig2}(b) and \ref{fig2}(e). The bands with predominantly $d_{xy}$ character have zero Chern number and are shown in Fig. \ref{fig2}(c,f). The bands with non-trivial Chern numbers are reported in Fig. \ref{fig2}(a,b,d,e); all of them exhibit the largest contribution to the Berry curvature along the lines $k_x$=$\pm{k_y}$, which are mirror lines when $h_z=0$. Turning on a finite $h_z$, we gap out the Dirac points present there for $h_z=0$ and produce avoided band crossings.
These crossings contribute to AHE as very localized positive/negative spikes of Berry curvature in the $k$-space.

\vskip 0.3cm
\textbf{Anomalous Hall conductance humps driven by orbital Rashba}
\vskip 0.2cm

\begin{figure}
\includegraphics[width=0.9\columnwidth]{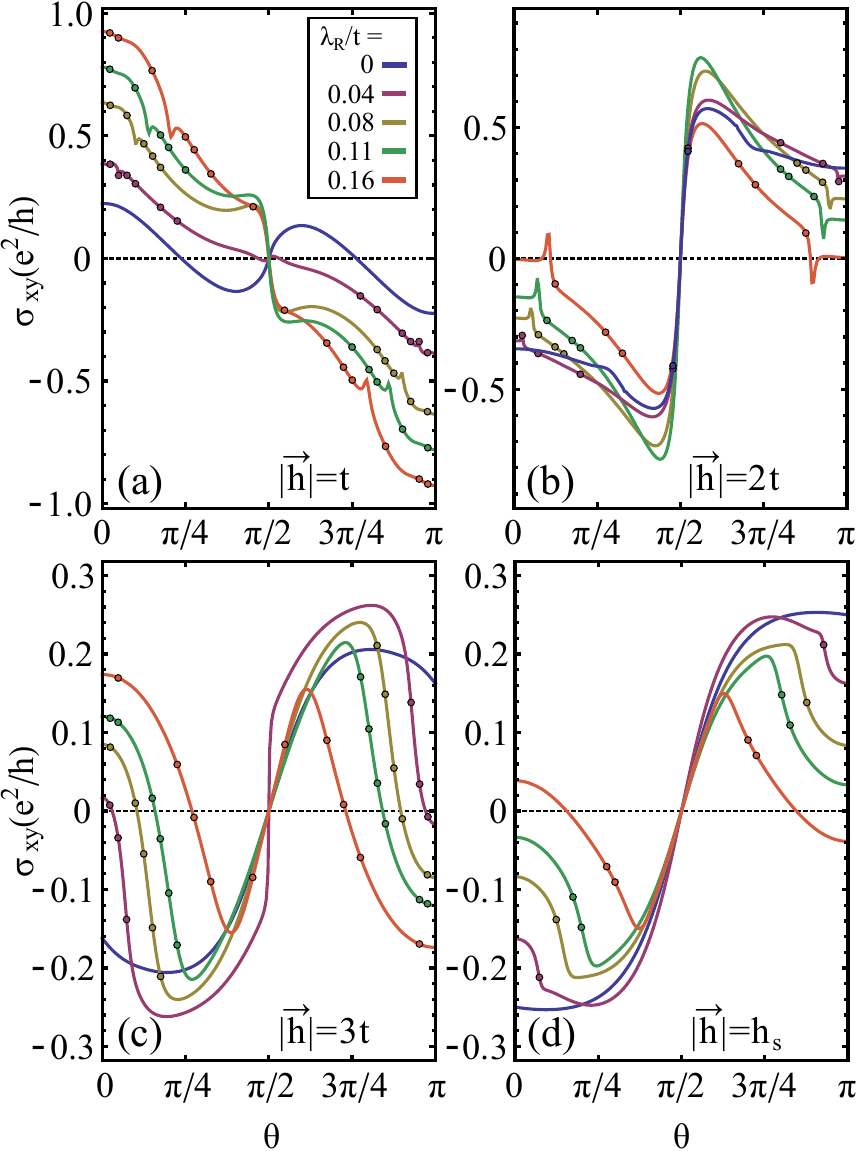}
\caption{Anomalous Hall conductivity $\sigma_{xy}$ as a function of polar angle
$\theta$ for effective Zeeman field $\vec{h}=|\vec{h}|\left(\sin\theta,0,\cos\theta\right)$
and for: (a) $|\vec{h}|=t$, (b) $|\vec{h}|=2t$, (c) $|\vec{h}|=3t$
and (d) $|\vec{h}|=h_{s}$, where at $h_{s}$ magnetization saturates
at the maximal value. 
The colored dots indicate the reshuffling of the Chern numbers of the bands.
The parameters are $t_{d}=-0.16t$, $\lambda=-0.2t$
and $\lambda_{R}/t=0,\,0.04,\,0.08,\,0.11,\,0.16$. The temperature
is $T=0.01t$. \label{fig3}}
\end{figure}

In this subsection, we calculate the  $\sigma_{xy}$ as a function of the polar angle $\theta$ between the magnetization and the z-axis. After completing this task, we can convert our results in the experimentally measured $\sigma_{xy}$ as a function of the magnetic field in the hysteretic process. 
The value of $\sigma_{xy}$ as a function of $\theta$ is reported in Fig. \ref{fig3}. We analyze it for different strengths of the magnetization $|\vec{h}|$ in panels (a,b,c,d). 
In a standard scenario where the Berry curvature remains constant with respect to $\theta$, we expect the  $\sigma_{xy}$ to follow the behaviour of M$_z$ and be proportional to $\cos(\theta)$, meaning that humps cannot be realized.
Through the magnetization direction, we can tune the energy of the Weyl points\cite{PhysRevResearch.1.032044,yao2018switchableweylnodestopological}, creating deep changes in the electronic structures. As a result, the anomalous Hall conductivity (AHC) for Weyl ferromagnets does not guarantee to have a standard behavior $\cos(\theta)$ as a function of the magnetization rotation. 
We note that without orbital Rashba and in the physically significant limit of large $|\vec{h}|$, $\sigma_{xy}$ tends to follow the $\cos(\theta)$. Increasing the orbital Rashba, we observe strong deviations from $\cos(\theta)$ in the limit of large $|\vec{h}|$. In particular, we can have a node in $\sigma_{xy}$ between 0 and $\frac{\pi}{2}$, when this happens we can have humps in the hysteresis loop if the value of $\sigma_{xy}$ is sufficiently small at $\theta$=0, as it will be clear in the rest of the paper.   

\begin{figure}
\includegraphics[width=1\columnwidth]{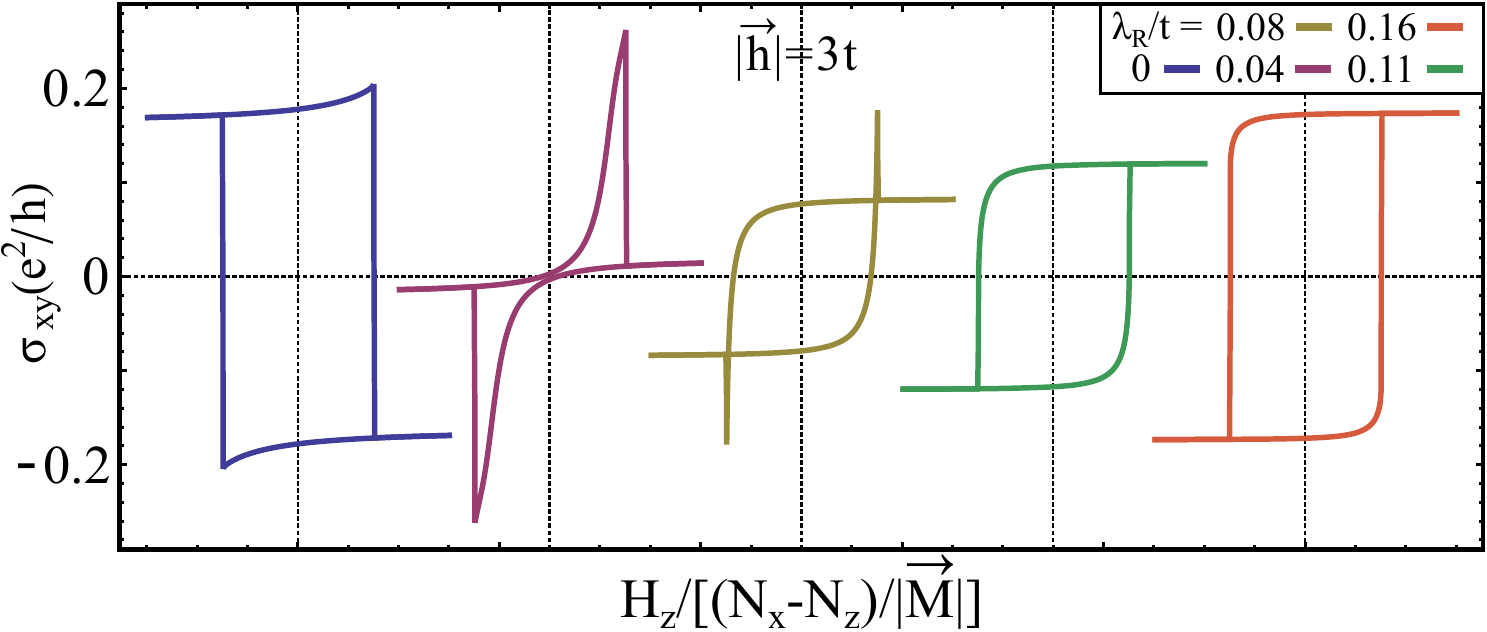}
\caption{Hysteresis loops of $\sigma_{xy}$ in external uniaxial magnetic field
$H_{z}$ obtained from the magnetic hysteresis
loop of the Stoner-Wolfarth model. Here we have assumed
that spin rotates in the $xz$ plane with the same angle
$\theta$ as the angle of the effective Zeeman field $\vec{h}=|\vec{h}|\left(\sin\theta,0,\cos\theta\right)$.
Dependence of $\sigma_{xy}$ on $\theta$ is taken from Fig. \ref{fig3}, while
$N_{x}$ $N_{z}$  are demagnetization factors along axes $x$ and $z$.
$\vec{M}$ is the spin magnetization. 
The parameters are $t_{d}=-0.16t$, $\lambda=-0.2t$ and $\lambda_{R}/t=0,\,0.04,\,0.08,\,0.11,\,0.16$.
 \label{fig4}}
\end{figure}

The ferromagnetic hysteresis is a non-equilibrium process where ferromagnetic domains tend to align with the external magnetic field. We consider what is happening in a single ferromagnetic domain and we map the rotation of spin magnetization on the hysteresis loop for the anomalous Hall conductivity $\sigma_{xy}$. This is done using the Stoner-Wohlfarth model as described in Appendix A, applying an external magnetic field defined as $H$. We report the results for the physically significant value of $|\vec{h}|$=3t in Fig. \ref{fig4}. Without orbital Rashba, we have a negative value of the AHC with a standard behavior as a function of $\theta$. Increasing the orbital Rashba to 0.04 and 0.08, the value of AHC at $\theta=0$ is small and positive, with the curve that presents a node as a function of $\theta$. If we apply a negative magnetic field, the spin rotates and $\theta$ increases, reaching a large negative value and creating the humps. By increasing the value of the orbital Rashba, we reach a limit where the AHC is large at $\theta$=0 such that the ordinary anomalous Hall effect covers the effect of the humps.

\begin{figure}
\includegraphics[width=1\columnwidth]{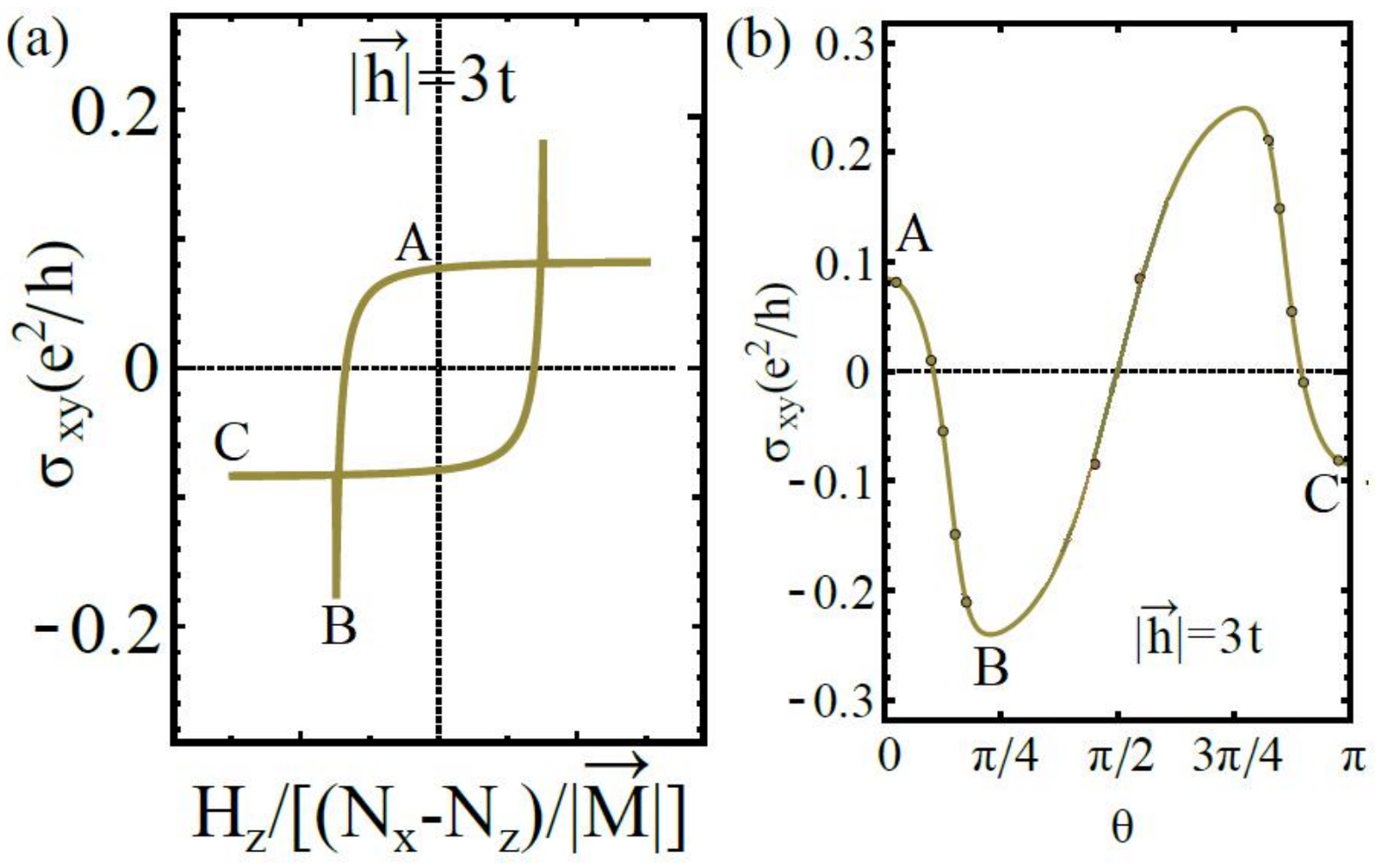}
\caption{(a) Hysteresis loops of $\sigma_{xy}$ in external uniaxial magnetic field $H_{z}$ and (b) Anomalous Hall conductivity $\sigma_{xy}$ as a function of polar angle $\theta$. Data for $|\vec{h}|$=3t and $\lambda_{R}/t$=0.08. A, B and C are the states of the 2D topological ferromagnets. A and C are extremely stable states, while state B is an intermediate state during the hysteresis loop.}\label{fig5}
\end{figure}

\vskip 0.3cm
\textbf{\Large Discussion and Conclusions}
\vskip 0.3cm

A simplified picture of how the humps are originated is reported in Fig. \ref{fig5} for the case of $|\vec{h}|$=3t and $\lambda_{R}/t$=0.08. 
At zero field, the system is in the stable state A with a positive Berry curvature. The stable state C is reached after the rotation of the spin by $\pi$.  The Berry curvature at the point C is equal in value but of the opposite sign to that at point A. The intermediate state B (or close to B) is stabilized in the Stoner-Wohlfarth model. The intermediate state B can generate humps in the AHC hysteresis loop only if the absolute value of the AHC at B is significantly greater than that at C. From our data, to obtain the value of AHC in B lower than C, we need the orbital Rashba coupling. The humps derive from spins that are not aligned with the easy axis. Humps are not expected at half-filling but are more likely to appear at intermediate filling levels (see Appendix A). 
In the saturation regime for the magnetization corresponding with |h|=h$_s$, which is probably the closest to the experimental situation, for $\lambda_R$=0, the AHE in Figure \ref{fig3}(d) is a cosine of the $\theta$ angle and cannot generate any hump. Therefore, the orbital Rashba is necessary to generate humps in the saturation limit. For lower values of |h|, we note that the breaking of the inversion symmetry is not strictly necessary for the appearance of the humps; However, this generates additional anticrossings, which serve as a source of Berry curvature, and strongly promotes oscillations in the AHE along with the emergence of hump-like features as a consequence of our model. The enhancement of the AHE due to inversion-symmetry breaking is supported by observations in systems such as noncentrosymmetric ferromagnets (e.g., GdPtBi\cite{Suzuki2016} TbPdBi\cite{Zhu2023}, and Co$_3$Sn$_2$S$_2$ derivatives\cite{Liu2018}), where inversion-symmetry breaking frequently coincides with unusually large AHE values.

In conclusion, we proposed a new microscopic model to describe the competing sources of Berry curvature in 2D topological ferromagnets, which can be extended to Weyl ferromagnets. When the absolute value of the anomalous Hall conductivity at zero field is close to zero, we can obtain humps in the anomalous Hall conductivity of Weyl ferromagnetic systems with breaking of inversion symmetry and orbital Rashba, but without invoking the Dzyaloshinskii-Moriya interaction (DMI) and scalar spin chirality.
In the presence of spin scalar chirality, additional contributions will be made to the anomalous Hall conductivity. Following our results, the size and robustness of the humps in the AHC depend on the number of spins not aligned to the easy axis. This agrees with the experimental observation that humps are observed over a wider range of magnetic field values when the field is not aligned with the z-axis\cite{Esser2021angular}. Therefore, boundaries of magnetic domains and impurities could enhance the effect of the humps.

The compound SrRuO$_3$, which could be a material's case to which this theory can be applied, can exhibit more effects than the ones reported here. 
For instance, as a future step for this investigation of SrRuO$_3$, we can mention the introduction of the staggered DMI\cite{PhysRevB.111.054442} that was neglected here. The staggered DMI produces spin cantings in the presence of both antiferromagnetic and ferromagnetic coupling; in the case of antiferromagnetic coupling, the systems are Weyl altermagnets\cite{li2024topologicalweylaltermagnetismcrsb,PhysRevB.110.085145} and the staggered DMI can produce weak ferromagnetism. In the case of ferromagnet, the staggered DMI can produce canting of the spins that will affect the description of the humps in ferromagnetic Weyl semimetals\cite{Shen2023-hr}.
Our results highlight the need for further investigation into systems exhibiting humps in anomalous Hall conductivities, as these features may arise from effects beyond spin scalar chirality. 
\\
Although the humps may reflect important signatures of the AHE's origin, disentangling the various contributions -- 
whether intrinsic or extrinsic -- to the AHE continues to be a considerable challenge \cite{Kimbell2022}. 
This understanding is relevant for the development of quantum electronic devices that 
utilize anomalous Hall and Berry curvature effects \cite{Cuoco2022}. 
\\
Finally, our study emphasizes the crucial role of orbital degrees of freedom and the underlying SU(3) structure of the Berry curvature in generating electronic bands with nontrivial topological characteristics. This framework leads to robust and competing Berry curvature contributions with opposite signs. We argue that, as for the case of oxide interfaces \cite{Lesne2023,Mercaldo2023}, in addition to unconventional Hall effects that are marked by magnetism and breaking of inversion symmetry, other nonstandard phenomena may also emerge within the nonlinear Hall regime \cite{Lesne2023}.

\vskip 0.3cm
\textbf{\Large Methods}
\vskip 0.3cm

\textbf{Berry curvatures for model Hamiltonian described by $2 \times 2$ and $3 \times 3$ matrices}
\vskip 0.2cm

A traceless 2D $2\times2$ $k-$space Hamiltonian can be written as,
\begin{equation}
{\cal H}_{{\bf k}}=\sum_{\alpha=1}^{3}b_{{\bf k}}^{\alpha}\sigma_{\alpha},
\end{equation}
where $\sigma_{\alpha}$ are the Pauli matrices. The two bands are given
by $E_{{\bf k}}^{\pm}=\pm|\vec{b}_{{\bf k}}|$. One can show that the Berry curvatures $\Omega$ (see \cite{Bar12}) for these bands are opposite and are
given by 
\begin{equation}
\Omega_{\pm}=\mp\frac{1}{2|\vec{b}_{{\bf k}}|^{3}}\vec{b}_{{\bf k}}\cdot\left[\partial_{1}\vec{b}_{{\bf k}}\times\partial_{2}\vec{b}_{{\bf k}}\right].
\end{equation}
To calculate the anomalous contribution to the Hall conductance we
need to integrate $\Omega_{\pm}$ up to the Fermi level.

Similarly, for a $3\times3$ one can also derive a closed-form Berry
curvature for the three bands \cite{Bar12}. A generic traceless $3\times3$ $k-$space
Hamiltonian can be written as,
\begin{equation}
{\cal H}_{{\bf k}}=\sum_{\alpha=1}^{8}b_{{\bf k}}^{\alpha}\lambda_{\alpha},
\end{equation}
where $\lambda_{\alpha}$ are the $8$ Gell-Mann matrices defined
as,
\begin{eqnarray}
\lambda_{1}=\begin{pmatrix}0 & 1 & 0\\
1 & 0 & 0\\
0 & 0 & 0
\end{pmatrix}, & \lambda_{2}=\begin{pmatrix}0 & -i & 0\\
i & 0 & 0\\
0 & 0 & 0
\end{pmatrix},\\
\lambda_{3}=\begin{pmatrix}1 & 0 & 0\\
0 & -1 & 0\\
0 & 0 & 0
\end{pmatrix}, & \lambda_{4}=\begin{pmatrix}0 & 0 & 1\\
0 & 0 & 0\\
1 & 0 & 0
\end{pmatrix},\nonumber \\
\lambda_{5}=\begin{pmatrix}0 & 0 & -i\\
0 & 0 & 0\\
i & 0 & 0
\end{pmatrix}, & \lambda_{6}=\begin{pmatrix}0 & 0 & 0\\
0 & 0 & 1\\
0 & 1 & 0
\end{pmatrix},\nonumber \\
\lambda_{7}=\begin{pmatrix}0 & 0 & 0\\
0 & 0 & -i\\
0 & i & 0
\end{pmatrix}, & \quad\lambda_{8}=\begin{pmatrix}\tfrac{1}{\sqrt{3}} & 0 & 0\\
0 & \tfrac{1}{\sqrt{3}} & 0\\
0 & 0 & \tfrac{-2}{\sqrt{3}}
\end{pmatrix}.\nonumber 
\end{eqnarray}
The bands are given by $E_{{\bf k}}^{n=1,2,3}$ in increasing order
and it can be shown that the Berry curvature of these bands is given
by
\begin{eqnarray}
\Omega_{n} & \!=\! & -\frac{4\xi_{n}^{3}}{3^{3/2}}\left(\gamma_{n}\vec{b}_{{\bf k}}\!+\!\vec{b}_{{\bf k}}*\vec{b}_{{\bf k}}\right)\\
 & \!\cdot\! & \left[\gamma_{n}^{2}\partial_{1}\vec{b}_{{\bf k}}\!\times\!\partial_{2}\vec{b}_{{\bf k}}\!+\!\gamma_{n}\partial_{1}\vec{b}_{{\bf k}}\!\times\!\partial_{2}\left(\vec{b}_{{\bf k}}*\vec{b}_{{\bf k}}\right)\right.\nonumber \\
 & \!+\! & \gamma_{n}\partial_{1}\left(\vec{b}_{{\bf k}}*\vec{b}_{{\bf k}}\right)\!\times\!\partial_{2}\vec{b}_{{\bf k}}\nonumber \\
 & \!+\! & \left.\partial_{1}\left(\vec{b}_{{\bf k}}*\vec{b}_{{\bf k}}\right)\!\times\!\partial_{2}\left(\vec{b}_{{\bf k}}*\vec{b}_{{\bf k}}\right)\right],\nonumber 
\end{eqnarray}
with parameters $\theta$, $\gamma$ and $\xi$ defined as
\begin{eqnarray}
\theta_{{\bf k}} & = & \tfrac{1}{3}\arccos\left[\frac{\vec{b}_{{\bf k}}\cdot\left(\vec{b}_{{\bf k}}*\vec{b}_{{\bf k}}\right)}{|\vec{b}_{{\bf k}}|^{3}}\right],\\
\gamma_{n} & = & 2|\vec{b}_{{\bf k}}|\cos\left(\theta_{{\bf k}}+\tfrac{2\pi}{3}n\right),\nonumber \\
\xi_{n} & = & \frac{1}{|\vec{b}_{{\bf k}}|^{2}\left[4\cos^{2}\left(\theta_{{\bf k}}+\tfrac{2\pi}{3}n\right)-1\right]}.\nonumber 
\end{eqnarray}
These formulas make use of the three kinds of vector-vector products
in $8-$dimensional space; ($i$) the standard scalar product, $\vec{a}\cdot\vec{b}=\sum_{i=1}^{8}a_{i}b_{i}$,
($ii$) antisymmetric vector product, $\left(\vec{a}\times\vec{b}\right)_{i}=\sum_{j,k=1}^{8}f_{ijk}a_{j}b_{k}$
and ($iii$) the symmetric vector product, $\left(\vec{a}*\vec{b}\right)_{i}=\sum_{j,k=1}^{8}d_{ijk}a_{j}b_{k}$.
For the $\times-$product we use the completely antisymmetric SU($3$)
structure factor $f_{ijk}$ given by,
\begin{eqnarray}
f_{123} & \!=\! & 1\\
f_{147} & \!=\! & -f_{156}\!=\!f_{246}\!=\!f_{257}\!=\!f_{345}\!=\!-f_{367}\!=\!\tfrac{1}{2}\nonumber \\
f_{458} & \!=\! & f_{678}=\tfrac{\sqrt{3}}{2}\nonumber 
\end{eqnarray}
where all other $f_{ijk}$ not related by a permutation of the above
indices vanish. Similarly, the completely symmetric SU($3$)
structure factor $d_{ijk}$, used for the $*-$product, is defined
by the equations,
\begin{eqnarray}
d_{118} & \!=\! & d_{228}\!=\!d_{338}\!=\!-d_{888}\!=\!\tfrac{1}{\sqrt{3}},\\
d_{448} & \!=\! & d_{558}\!=\!d_{668}\!=\!d_{778}\!=\!-\tfrac{1}{2\sqrt{3}},\nonumber \\
d_{146} & \!=\! & d_{157}\!=\!-d_{247}\!=\!d_{256}\!=\!\tfrac{1}{2},\nonumber \\
d_{344} & \!=\! & d_{355}\!=\!-d_{366}\!=\!-d_{377}\!=\!\tfrac{1}{2}.\nonumber 
\end{eqnarray}
These structure factors are useful to define the multiplication law for Gell-Mann matrices, in a similar way as the Levi-Civita symbol defines the multiplication of the Pauli matrices; we have
\begin{equation}
\lambda_{\alpha}\lambda_{\beta}=\tfrac{2}{3}\delta_{\alpha\beta}+\sum_{\gamma=1}^{8}\left(d_{\alpha\beta\gamma}+if_{\alpha\beta\gamma}\right)\lambda_{\gamma}.
\label{mult}
\end{equation}

\noindent Finally, the energy levels for the $3\times3$ case can be obtained
from the eigenstate projector $P_{n}$ \cite{Bar12}. We have
\begin{equation}
P_{n}=\tfrac{1}{3}\left(1+\sqrt{3}\sum_{\alpha=1}^{8}n_{{\bf k}}^{\alpha}\lambda_{\alpha}\right),
\end{equation}
with
\begin{equation}
\vec{n}_{{\bf k}}=\xi_{n}\left(\gamma_{n}\vec{b}_{{\bf k}}+\vec{b}_{{\bf k}}*\vec{b}_{{\bf k}}\right).
\end{equation}
Then the energy bands are given by
\begin{equation}
E_{{\bf k}}^{n}={\rm Tr}\left(P_{n}{\cal H}_{{\bf k}}\right),
\end{equation}
which can be simplified using the multiplication rule (\ref{mult}) and tracelessness of the Gell-Mann matrices to 
\begin{equation}
E_{{\bf k}}^{n}=\tfrac{2\sqrt{3}}{3}\xi_{n}\left(\gamma_{n}\vec{b}_{{\bf k}}+\vec{b}_{{\bf k}}*\vec{b}_{{\bf k}}\right)\cdot b_{{\bf k}}.
\end{equation}

\begin{figure}
\includegraphics[width=1\columnwidth]{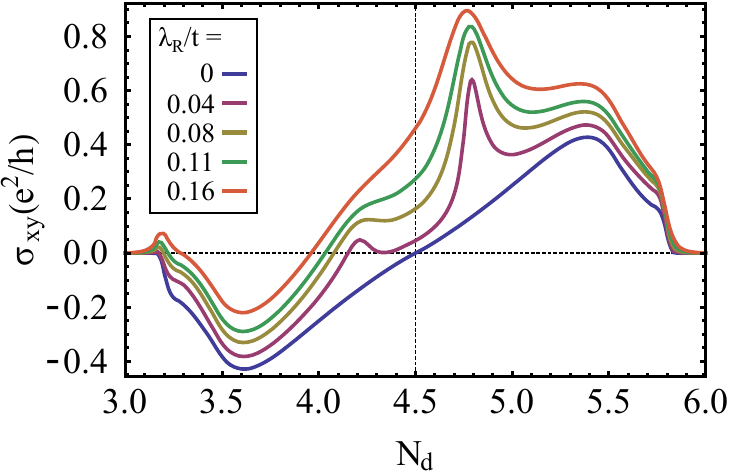}\caption{Anomalous Hall conductivity $\sigma_{xy}$ as a function of filling
factor of the $t_{2g}$ $d$-shell $N_{d}$. The parameters are $t_{d}=-0.16t$,
$\lambda=-0.2t$ and $\lambda_{R}/t=0,\,0.04,\,0.08,\,0.11,\,0.16$
for decreasing global minimum $\sigma_{xy}$, respectively. The temperature
is $T=0.01t$ and the applied field corresponds to the saturation amplitude. \label{fig1s}}
\end{figure}

\begin{figure*}
\includegraphics[width=1\textwidth]{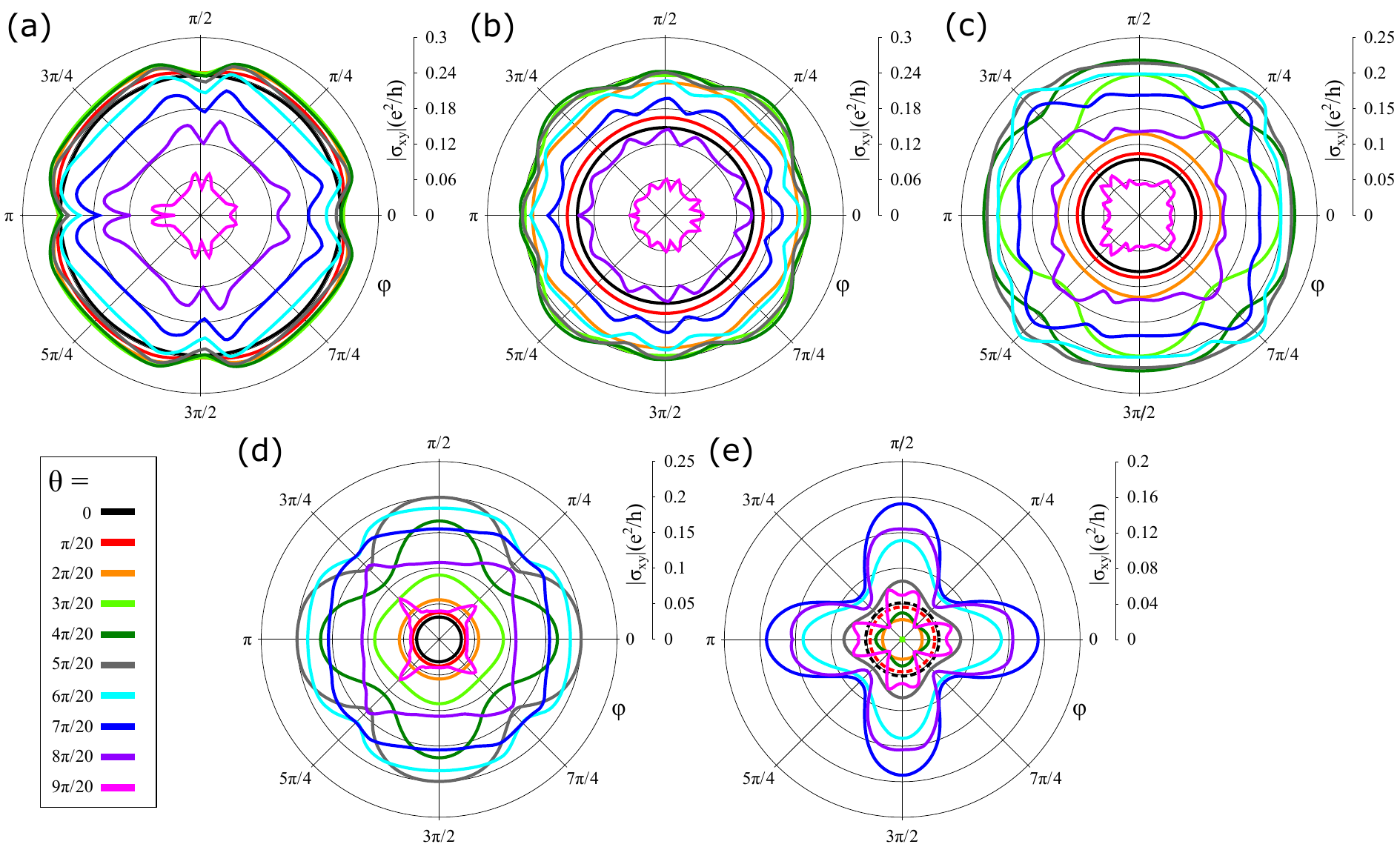}\caption{High-field anomalous Hall conductivity $\sigma_{xy}$ as a function
of field directions given by polar and azimuthal angles $(\theta,\phi)$
for different values of $\lambda_{R}/t=0,\,0.04,\,0.08,\,0.11,\,0.16$
(a-e). The parameters are $t_{d}=-0.16t$ and $\lambda=-0.2t$. The
solid (dashed) lines indicate that $\sigma_{xy}$ is negative (positive). $\sigma_{xy}$
vanishes for $\theta=\pi/2$ and any $\phi$ and satisfies $\sigma_{xy}(\pi/2+\theta,\phi)=-\sigma_{xy}(\pi/2-\theta,\phi)$.
The temperature is set to $T=0.1t$.\label{fig2s}}
\end{figure*}

\vskip 0.3cm
\textbf{Appendix A: Electron filling and angular dependence of $\sigma_{xy}$}
\vskip 0.2cm

In this Appendix, we report the dependence on $\sigma_{xy}$ from the filling of d-electrons N$_d$ and its angular distribution. 

The system under investigation has 6 d-electrons; we aim to analyze the $\sigma_{xy}$ and relative humps as a function of the filling of d-electrons. The properties of the system are symmetric with respect to half-filling, so we will report just the results for the filling from 3 to 6.  
The values of $\sigma_{xy}$ as a function of the filling N$_d$ for different orbital Rashba $\lambda_{R}/t$ are reported in 
Fig. \ref{fig1s}.  
Without orbital-Rashba, $\sigma_{xy}$ is antisymmetric with respect to the filling N$_d$=4.5, which agrees with other models for metallic systems hosting with nontrivial Chern number \cite{Kang2022xq}. The AHC in Fig. \ref{fig1s} is linear with no anomalies in the absence of orbital Rashba. In the presence of orbital Rashba, we report humps in the $\sigma_{xy}$ also as a function of the d-electron filling N$_d$, confirming that the orbital-Rashba effect is one of the key ingredients in the origin of the humps.
In particular, fillings between 4 and 5 are more suitable for the creation of humps. The orbital Rashba is ineffective at half-filling (N$_d$=3) and full filling (N$_d$=6). Therefore, no humps are expected at half filling.
The non-linear behaviour of $\sigma_{xy}$ calculated with orbital-Rashba shows similarities with $\sigma_{xy}$ calculated in the presence of skyrmions\cite{Verma2022-gt}.

In Fig. \ref{fig2s}, we present the angular distribution of the $\sigma_{xy}$ as a function of the azimuthal angle $\phi$. The 5 plots in Fig. \ref{fig2s}(a-e) refer to different values of the orbital Rashba. Inside every panel, we plot 10 different values of the polar angle $\theta$. We can observe that the system hosts a C$_4$ symmetry. While the system is weakly anisotropic for low values of $\lambda_{R}/t$, it becomes strongly anisotropic for large values of $\lambda_{R}/t$. At the largest value $\lambda_{R}/t$=0.16, $\sigma_{xy}$ exhibits lobes resembling those of the real spherical harmonic function x$^2$-y$^2$.

\vskip 0.3cm
\textbf{Appendix B: Stoner-Wohlfarth model to obtain $\rho_{xy}$ as a function of the external magnetic field}
\vskip 0.2cm

For the single-domain ferromagnet, the Stoner-Wohlfarth allows us to obtain the magnetization loop once we know the module of the magnetization $|\vec{M}|$, the magnetocrystalline anisotropy K$_2$, the module $H_Z$ and the direction of the applied magnetic field $\theta_H$, polar angle of the magnetic field relative to the easy axis.
The total energy of the system is described by:
\begin{equation}
E(\theta)=K_2\cos^2(\theta)-|\vec{M}|H_Z\cos(\theta-\theta_H).
\end{equation}
where $\theta$ is the orientation of the magnetization domain that changes during the hysteresis loop.
We minimize the energy functional as a function of $\theta$, and we obtain the magnetization along the z-axis, which is achieved by projecting the magnetization vector along the z-axis using $ M_z = |\vec{M}|\cos(\theta)$. 
When $\theta_H=0$, the magnetic field is along the easy axis and two minima of the total energy are achieved just for the two values $\theta=0,\pi$. The hysteresis loop takes the form of a simple square loop with coercive field  $\frac{2K_2}{M_S}$. 
When $\theta_H$ differs from zero, the minimum of the energy is achieved for the whole interval of the value of $\theta$ from 0 to $\pi$ with a smoothing of the magnetization loop in the region of the flipping of the magnetization and a reduction of the coercive field\cite{Autieri2014NJP}. 
When $\theta_H$=$\pi$, the field is along the hard axis, and there is no hysteresis since the coercive field is zero. Since we have a single-domain ferromagnet, we could use the Stoner-Wohlfarth model not only to obtain the magnetization loop but also to obtain the anomalous Hall conductivity loop $\rho_{xy}$.
We use as 
$\rho_{xy}$=$\rho_{xy}(\theta)$ as shown in Fig. \ref{fig3}
of the main text.
Suitable numerical values for the SrRuO$_3$ are
M$_S$=1.5 $\mu_B$ and K$_2$=$\frac{5}{2}|\vec{M}|$ Tesla
as obtained from the literature\cite{Ziese2010structural}, while we have used a small value of $\theta_H$ to obtain $\rho_{xy}$ in the whole range of $\theta$. The Stoner-Wohlfarth model has been recently used to describe the magnetization loop of Ru-oxides; therefore, it is suitable for this task \cite{Piamonteze2021}.

\vskip 0.5cm
\textbf{\Large Data Availability}\\
\vskip 0.1cm
The data that support the findings of this study are available from the corresponding author upon reasonable request.

\vskip 0.5cm
\textbf{\Large Code Availability}\\
\vskip 0.1cm
The code that supports the findings of this study is available from the corresponding author upon reasonable request.

\vskip 0.5cm
\textbf{\Large Author Contribution}\\
\vskip 0.1cm
W. B. developed the computational models and performed the numerical simulations and data analysis. C. A. wrote the initial draft of the manuscript with input from all authors. M. C. supervised the research project. All authors contributed extensively to the discussion, the conception and the analysis of the results presented in this paper. 

\vskip 0.5cm
\textbf{\Large Acknowledgements}\\
\vskip 0.1cm
C.A. acknowledges X. Gong for useful discussions.
This research was supported by the "MagTop" project (FENG.02.01-IP.05-0028/23) carried out within the "International Research
Agendas" programme of the Foundation for Polish Science, co-financed by the
European Union under the European Funds for Smart Economy 2021-2027 (FENG). W.B. also acknowledges support by Narodowe Centrum Nauki (NCN, National Science Centre, Poland) Project No. 2019/34/E/ST3/00404.
C.A. and M.C. acknowledge support from PNRR MUR project PE0000023-NQSTI. M.C. acknowledges support by Italian Ministry of University and Research (MUR) PRIN 2022 under the Grant No. 2022LP5K7 (BEAT).

\vskip 0.5cm    
\textbf{\Large Competing interests}\\
\vskip 0.1cm
The Authors declare no Competing Financial or Non-Financial Interests.

\vskip 0.5cm
\textbf{\Large Additional information}\\
\vskip 0.1cm
Correspondence should be addressed to Mario Cuoco, email: mario.cuoco@spin.cnr.it, Wojciech Brzezicki, email: w.brzezicki@uj.edu.pl and Carmine Autieri, email: autieri@magtop.ifpan.edu.pl.

\bibliography{CRO_EF}

\end{document}